\documentclass[showpacs,pre,superscriptaddress,twocolumn]{revtex4}%
\usepackage{amssymb}
\usepackage{graphicx}
\usepackage{amsmath}
\usepackage{amsfonts}%
\begin{document}
\title{Quantum games of asymmetric information}
\author{Jiangfeng Du}
\email{djf@ustc.edu.cn}
\affiliation{Structure Research Laboratory and Department of Modern Physics, University
of Science and Technology of China, Hefei, 230027, People's Republic of China}
\affiliation{Department of Physics, National University of Singapore, Lower Fent Ridge,
Singapore 119260, Singapore}
\affiliation{Centre for Quantum Computation, Department of Applied Mathematics and
Theoretical Physics, University of Cambridge, Wilberforce Road, Cambridge
CB3 0WA, United Kingdom}
\author{Hui Li}
\email{lhuy@mail.ustc.edu.cn}
\affiliation{Structure Research Laboratory and Department of Modern Physics, University
of Science and Technology of China, Hefei, 230027, People's Republic of China}
\author{Chenyong Ju}
\affiliation{Structure Research Laboratory and Department of Modern Physics, University
of Science and Technology of China, Hefei, 230027, People's Republic of China}

\begin{abstract}
We investigate quantum games in which the information is asymmetrically
distributed among the players, and find the possibility of the quantum game
outperforming its classical counterpart depends strongly on not only the
entanglement, but also the informational asymmetry. What is more interesting,
when the information distribution is asymmetric, the contradictive impact of
the quantum entanglement on the profits is observed, which is not reported in
quantum games of symmetric information.

\end{abstract}
\pacs{02.50.Le, 03.67.-a}
\maketitle

\section{Introduction}

The field of information and computation has experienced a fundamental
innovation since the last decades of the twentieth century, through the
combination with the theory of quantum physics. The new-born theory of quantum
information and computation opens a broad field of potential applications
\cite{1}. Its recent application into the theory of games extends the
classical game theory \cite{8}, which is in fact one of the cornerstones of
modern economics, into the quantum domain. It has shown that quantum games may
have great advantages over their classical counterparts
\cite{2,3,4,6,g1,g2,g3,g4}. Many of the current works focus on games in which
the players has finite number of classical strategies and/or the information
is symmetrically distributed among the players. While games with continuous
set of strategies and those of asymmetric information, which represent much
realistic significance \cite{5}, especially in market situations in economics,
are not given much attention. However the quantization of these games deserves
thorough investigations, and interesting results could be obtained.

The investigations on quantum games might provide new insights into the field
of economics research, as it does in the field of computation, communication
and others. There are several reasons that quantizing games that could be
applied in economics may be interesting. First, market situations could be in
their nature regarded as games, their quantization may be of the same
interests as quantizing games \cite{3}. Second, in any market situation,
information and communication are of most importance. However, information is
--- as we live in a quantum world --- legitimate to think of as quantum
information, and communication may also need to think of as quantum
communication (at least in the near future) \cite{1}, it therefore might be
interesting to investigate the quantization of market situations as games, and
interesting quantum features might be explored.

In this paper we investigate the quantum form of a particular game of the
market situation known as the Cournot's Duopoly \cite{7} of asymmetric
information, based on the previously proposed physical model for
continuous-variable quantum games \cite{6}. In the quantum game of asymmetric
information, the \textquotedblleft interaction\textquotedblright\ between the
quantum entanglement and the informational asymmetry creates interesting
properties of the game. Due to the presence of informational asymmetry, the
quantum entanglement has contradictive effects: on the one hand it promotes
cooperation and potentially increases the profits, but on the other hand it
potentially decreases the profits at the same time. Whether the quantum game
outperforms its classical counterpart depends strongly on not only the quantum
entanglement but also the informational asymmetry.

\section{Classical Cournot's Duopoly of asymmetric information}

We now briefly recall the classical Cournot's Duopoly \cite{7} of asymmetric
information. In a simple scenario, firm 1 and firm 2 simultaneously choose the
quantities (the strategies) $q_{1}$ and $q_{2}$, respectively, of a
homogeneous product. Let $Q=q_{1}+q_{2}$ be the total quantity, and the market
price be%
\begin{equation}
P\left(  Q\right)  =\left\{
\begin{array}
[c]{ll}%
a-Q & \text{, for }Q\leqslant a\\
0 & \text{, for }Q>a
\end{array}
\right.  .
\end{equation}
Denote the unit cost of firm 1 and 2 by $c_{1}$ and $c_{2}$ respectively, with
$c_{j}<a$ ($j=1,2$). Then the profit for firm $j$ is%
\begin{equation}
u_{j}\left(  q_{1},q_{2}\right)  =q_{j}\left[  P\left(  Q\right)
-c_{j}\right]  ,
\end{equation}
with $j=1,2$. In the case of asymmetric information, firm 1 does not clearly
know what $c_{2}$ (firm 2's unit cost) is, only knows that $c_{2}=c_{H}$ with
probability $\theta$ and $c_{2}=c_{L}$ with probability $1-\theta$
($c_{H}>c_{L}$). Yet firm 2 knows with certainty the unit cost of its product
($c_{2}$) as well as that of firm 1's ($c_{1}$). Let $q_{2H}^{\ast}$ and
$q_{2L}^{\ast}$ be the quantity of firm 2 when $c_{2}=c_{H}$ and $c_{2}=c_{L}
$, respectively, and $q_{1}^{\ast}$ be the quantity of firm 1. If
$c_{2}=c_{H\left(  L\right)  }$, then firm 2 needs to set $q_{2}=q_{2H\left(
L\right)  }^{\ast}$ to maximize its profit%
\begin{equation}
u_{2H\left(  L\right)  }\left(  q_{1}^{\ast},q_{2}\right)  =q_{2}%
[(a-q_{1}^{\ast}-q_{2})-c_{H\left(  L\right)  }].
\end{equation}
Firm 1 needs to set $q_{1}=q_{1}^{\ast}$ to maximize its \textit{expected}
profit%
\begin{equation}
u_{1}\left(  q_{1},q_{2H}^{\ast},q_{2L}^{\ast}\right)  =\theta u_{1}\left(
q_{1},q_{2H}^{\ast}\right)  +(1-\theta)u_{1}\left(  q_{1},q_{2L}^{\ast
}\right)  ,
\end{equation}
where%
\begin{equation}
u_{1}(q_{1},q_{2})=q_{1}[(a-q_{1}-q_{2})-c_{1}].
\end{equation}
Solving the three optimization problems yields the Bayes-Nash equilibrium
\cite{8}%

\begin{align}
q_{1}^{\ast} &  =\frac{2k_{1}-k_{2}}{3},\nonumber\\
q_{2H}^{\ast} &  =\frac{a+c_{1}-2c_{H}}{3}+\frac{(1-\theta)\Delta}%
{6},\nonumber\\
q_{2L}^{\ast} &  =\frac{a+c_{1}-2c_{L}}{3}-\frac{\theta\Delta}{6},
\end{align}
where%
\begin{align}
k_{1} &  =a-c_{1},\nonumber\\
k_{2} &  =a-[\theta c_{H}+(1-\theta)c_{L}],\nonumber\\
\Delta &  =c_{H}-c_{L}.\label{eq 5}%
\end{align}
The special instance with $k_{1}=k_{2}=k$ and $\Delta=0$ reduces to the
original model of symmetric information, with the unique Nash equilibrium%
\begin{equation}
q_{1}^{\ast}=q_{2}^{\ast}=\frac{k}{3},
\end{equation}
and the payoffs being%
\begin{equation}
u_{1}\left(  q_{1}^{\ast},q_{2}^{\ast}\right)  =u_{2}\left(  q_{1}^{\ast
},q_{2}^{\ast}\right)  =\frac{k^{2}}{9}.
\end{equation}
However this equilibrium fails to be the Pareto optimum \cite{2}, which could
easily be found to be%
\begin{equation}
q_{1}^{\prime}=q_{2}^{\prime}=\frac{k}{4},
\end{equation}
with%
\begin{equation}
u_{1}(q_{1}^{\prime},q_{2}^{\prime})=u_{2}(q_{1}^{\prime},q_{2}^{\prime
})=\frac{k^{2}}{8}.
\end{equation}

\section{Quantum Cournot's Duopoly of asymmetric information}

\begin{figure}
[ptb]
\begin{center}
\includegraphics[
height=0.9893in,
width=3.1955in
]%
{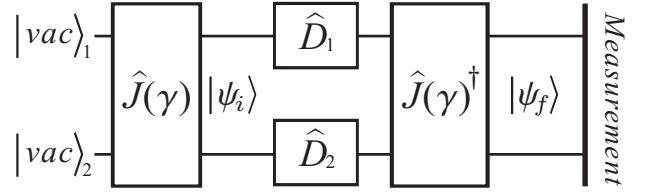}%
\caption{The quantum structure of the Cournot's Duopoly.}%
\label{Model}%
\end{center}
\end{figure}

The quantum structure is given in Fig. \ref{Model}, which is the same as
presented in Ref. \cite{6}. The necessity to include continuous-variable
quantum systems is that a continuous set of \textit{distinguishable }states
are necessary to represent all the possible outcomes of classical strategies,
due to the \textit{distinguishability} of classical strategies. In Fig.
\ref{Model}, $\left\vert vac\right\rangle _{1}$ and $\left\vert
vac\right\rangle _{2}$ are two vacuum states, \textit{e.g.} of two single-mode
electromagnetic fields respectively belong to the two firms. $\hat{J}\left(
\gamma\right)  $ and $\hat{J}\left(  \gamma\right)  ^{\dag}$ are unitary
operators, which are known to both firms and should be symmetric with respect
to the interchange of the two firms to guarantee a fair competition. The
initial state of the game is%
\begin{equation}
\left\vert \psi_{i}\right\rangle =\hat{J}\left(  \gamma\right)  \left\vert
vac\right\rangle _{1}\left\vert vac\right\rangle _{2}.
\end{equation}
Strategic moves of firm $j$ is associated with unitary local operator $\hat
{D}_{j}$. The final state of the game is denoted by%
\begin{equation}
\left\vert \psi_{f}\right\rangle =\hat{J}\left(  \gamma\right)  ^{\dagger
}(\hat{D}_{1}\otimes\hat{D}_{2})\hat{J}\left(  \gamma\right)  \left\vert
vac\right\rangle _{1}\left\vert vac\right\rangle _{2}.
\end{equation}
%

It is straightforward to set the final measurement be corresponding to the
observables $\hat{X}_{j}=(\hat{a}_{j}^{\dagger}+\hat{a}_{j})/\sqrt{2}$ (the
\textquotedblleft position\textquotedblright\ operators) for firm $j$, where
$\hat{a}_{j}^{\dagger}$ ($\hat{a}_{j}$) is the creation (annihilation)
operator of firm $j$'s electromagnetic field. If the measurement result is
$\tilde{x}_{j}$, then the individual quantity is determined by $q_{j}%
=\tilde{x}_{j}$, and hence the profit by%
\begin{equation}
u_{j}^{Q}(\hat{D}_{1},\hat{D}_{2})=u_{j}(\tilde{x}_{1},\tilde{x}_{2}),
\end{equation}
where the superscript \textquotedblleft$Q$\textquotedblright\ denotes
\textquotedblleft quantum\textquotedblright. However, as will be shown in Eq.
(\ref{eq f}), in the case we considered in the present paper, the final state
of the game $\left\vert \psi_{f}\right\rangle $ is a tensor product of two
coherent states, respectively belonging to the two firms. One can not have a
deterministic measurement result of $\hat{X}_{j}$, since a coherent state is
not an eigenstate of $\hat{X}_{j}$. This poses a problem because the quantity
$q_{j}$ is affected by uncertainty $\Delta q_{j}^{2}=\frac{1}{2} $. One
possible method to reduce this uncertainty is to perform appropriate
\textit{squeezing} operation on the final state before the measurement
according to $\hat{X}_{j}$ is carried out. The uncertainty of the measurement
result of $\hat{X}_{j}$ could be reduced, at the cost of increasing the
uncertainty of the measurement result of $\hat{P}_{j}$. In this paper, we
assume the limit case that the state is infinitely squeezed, so that the
uncertainty of the measurement result of $\hat{X}_{j}$ tends to zero.
Consequently, given a coherent state $\exp(-i\tilde{x}_{j}^{\prime}\hat{P}%
_{j})\left\vert vac\right\rangle _{j}$, the final measurement could
deterministically yield $q_{j}=\tilde{x}_{j}^{\prime}$ in this limit.

\begin{figure*}[ptb]
\begin{center}
\includegraphics[
height=2.2961in, width=5.9871in
]{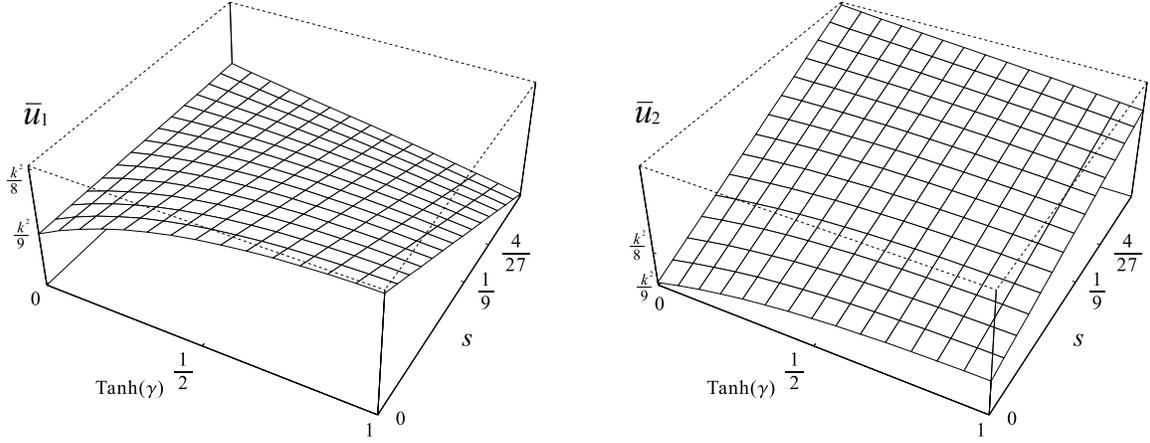}
\end{center}
\caption{The profits in the iterative game with $k_{1}=k_{2}=k$, with respect
to $\tanh(\gamma)$ and the amount of informational asymmetry $s$
($\tanh(\gamma)$ monotonously maps $\gamma\in\left[  0,\infty\right)  $ into
$\tanh(\gamma)\in\left[  0,1\right)  $). $\bar{u}_{1}$ is at the left and
$\bar{u}_{2}$\ at the right.}%
\label{Payoff}%
\end{figure*}

The classical Cournot's Duopoly can be faithfully represented when $\hat
{J}\left(  \gamma\right)  =\hat{J}\left(  \gamma\right)  ^{\dagger}=I$ (the
identity operator). The set%
\begin{equation}
S_{j}=\{\hat{D}_{j}\left(  x_{j}\right)  =\exp(-ix_{j}\hat{P}_{j})\mid
x_{j}\in\left[  0,\infty\right)  \}
\end{equation}
is the quantum counterpart of the classical strategic space, where $\hat
{P}_{j}=i(\hat{a}_{j}^{\dagger}-\hat{a}_{j})/\sqrt{2}$ (the \textquotedblleft
momentum\textquotedblright\ operators). In this paper, we restrict ourselves
to the \textquotedblleft minimal\textquotedblright\ extension, \textit{i.e.}
we maintain the strategic space unexpanded ($S_{j}$ for firm $j$) while only
extend the initial state $\left\vert \psi_{i}\right\rangle $ to be entangled.
This minimal extension guarantees that any features of the game not seen in
the classical form could be completely due to the quantum entanglement.
However it is also possible to find a quantum version that includes both the
entangled state and the expanded strategic spaces.

The choice of the entangling operator is not unique. Even though the
requirement that for vanishing entanglement the classical game should be
reproduced can not uniquely specify this operator in the case presented in
this paper. However a possible and legitimate one is%
\begin{equation}
\hat{J}\left(  \gamma\right)  =e^{-\gamma(\hat{a}_{1}^{\dagger}\hat{a}%
_{2}^{\dagger}-\hat{a}_{1}\hat{a}_{2})}=e^{i\gamma(\hat{X}_{1}\hat{P}_{2}%
+\hat{X}_{2}\hat{P}_{1})}.\label{eq 1}%
\end{equation}
The initial state is exactly the two-mode squeezed vacuum state%
\begin{equation}
\left\vert \psi_{i}\right\rangle =\exp\{-\gamma(\hat{a}_{1}^{\dagger}\hat
{a}_{2}^{\dagger}-\hat{a}_{1}\hat{a}_{2})\}\left\vert vac\right\rangle
_{1}\left\vert vac\right\rangle _{2},
\end{equation}
where $\gamma\geqslant0$ is known as the squeezing parameter and can be
reasonably regarded as a measure of entanglement. Detailed calculation reveals
that if firm $j$'s strategy is $\hat{D}_{j}\left(  x_{j}\right)  =\exp
(-ix_{j}\hat{P}_{j})$, then the final state is%
\begin{align}
\left\vert \psi_{f}\right\rangle = & \exp\{-i(x_{1}\cosh\gamma+x_{2}%
\sinh\gamma)\hat{P}_{1}\}\left\vert vac\right\rangle _{1}\nonumber\\
& \otimes\exp\{-i(x_{2}\cosh\gamma+x_{1}\sinh\gamma)\hat{P}_{2}\}\left\vert
vac\right\rangle _{2}.\label{eq f}%
\end{align}
Hence the quantities read out from the final measurement are%
\begin{align}
q_{1} &  =x_{1}\cosh\gamma+x_{2}\sinh\gamma,\nonumber\\
q_{2} &  =x_{2}\cosh\gamma+x_{1}\sinh\gamma.
\end{align}
The total quantity is $Q=q_{1}+q_{2}=e^{\gamma}\left(  x_{1}+x_{2}\right)  $,
and the market price is $P=a-e^{\gamma}\left(  x_{1}+x_{2}\right)  $.
Therefore profits are%
\begin{align}
u_{1}^{Q}\left(  x_{1},x_{2}\right)   &  =\left(  x_{1}\cosh\gamma+x_{2}%
\sinh\gamma\right)  \left[  P-c_{1}\right]  ,\nonumber\\
u_{2H\left(  L\right)  }^{Q}\left(  x_{1},x_{2}\right)   &  =\left(
x_{2}\cosh\gamma+x_{1}\sinh\gamma\right)  \left[  P-c_{H\left(  L\right)
}\right]  ,\label{eq 2}%
\end{align}
here for convenience we directly denote the strategy by $x_{j}$ when it is
$\hat{D}_{j}\left(  x_{j}\right)  $.

Let $\{x_{1}^{\ast},x_{2H}^{\ast},x_{2L}^{\ast}\}$ be the Bayes-Nash
equilibrium. Then $x_{2}=x_{2H\left(  L\right)  }^{\ast}$ is chosen to
maximize $u_{2H\left(  L\right)  }^{Q}(x_{1}^{\ast},x_{2})$, and $x_{1}%
=x_{1}^{\ast}$ is chosen to maximize $\theta u_{1}^{Q}\left(  x_{1}%
,x_{2H}^{\ast}\right)  +(1-\theta)$ $u_{1}^{Q}\left(  x_{1},x_{2L}^{\ast
}\right)  $. Solving the three optimization problem yields the Bayes-Nash
equilibrium \cite{8} and the profits could also be obtained. For convenience
and simplicity, we further set $k_{1}=k_{2}=k$. Detailed calculation gives the
unique Bayes-Nash equilibrium as%
\begin{align}
x_{1}^{\ast}  & =\frac{k\cosh\gamma}{1+2e^{2\gamma}},\nonumber\\
x_{2H}^{\ast}  & =\frac{k-\left(  1-\theta\right)  \Delta+e^{2\gamma}\left[
k-2\left(  1-\theta\right)  \Delta\right]  }{2e^{\gamma}\left(  1+2e^{2\gamma
}\right)  },\nonumber\\
x_{2L}^{\ast}  & =\frac{k+\theta\Delta+e^{2\gamma}\left[  k+2\theta
\Delta\right]  }{2e^{\gamma}\left(  1+2e^{2\gamma}\right)  }.
\end{align}

In the remaining part of this paper, we would like to consider an iterative
game in which the unit cost of firm 2's product is determined by the
probability known by firm 1, \textit{i.e.} $c_{H}$ with probability $\theta$
and $c_{L}$ with probability $1-\theta$, to avoid the ambiguity and complexity
caused by the specific choice of firm 2's unit cost in a single game. The
\textit{average} profits in the iterative game are%
\begin{align}
\bar{u}_{1}\left(  \gamma,s\right)   &  =\theta u_{1}^{Q}\left(  x_{1}^{\ast
},x_{2H}^{\ast}\right)  +\left(  1-\theta\right)  u_{1}^{Q}\left(  x_{1}%
^{\ast},x_{2L}^{\ast}\right) \nonumber\\
&  =\frac{k^{2}}{8}\left[  \frac{8e^{\gamma}\cosh\gamma}{(3\cosh\gamma
+\sinh\gamma)^{2}}+\left(  e^{-2\gamma}-1\right)  s\right]  ,\nonumber\\
\bar{u}_{2}\left(  \gamma,s\right)   &  =\theta u_{2H}^{Q}\left(  x_{1}^{\ast
},x_{2H}^{\ast}\right)  +\left(  1-\theta\right)  u_{2L}^{Q}\left(
x_{1}^{\ast},x_{2L}^{\ast}\right) \nonumber\\
&  =\bar{u}_{1}\left(  \gamma,s\right)  +\frac{k^{2}}{4}s,\label{eq 3}%
\end{align}
where%
\begin{equation}
s=\theta\left(  1-\theta\right)  \frac{\Delta^{2}}{k^{2}}\geqslant
0.\label{eq s}%
\end{equation}
The profits are already expressed as functions of $\gamma$ and $s$, and are
plotted in Fig. \ref{Payoff}.

The notation $s$ defined in Eq. (\ref{eq s}) can reasonably be regarded as the
\textit{amount of informational asymmetry}. Indeed, $s=0$ is attained only
when $\theta=0$, $\theta=1$, or $\Delta=0$, each corresponding to the case
where firm 1 has the perfect information about firm 2's unit cost,
\textit{i.e.} there is no asymmetry in the information distribution. However
for fixed $\theta$, $s$ increases as $\Delta$ increases, and for fixed
$\Delta$, $s$ increases as $\theta$ approaches $1/2$. This means that the more
asymmetrical the information distribution is, the larger is $s$. It is in this
sense that we regard $s$ as a measure of the informational asymmetry of the game.

We now investigate how the profits depend on the entanglement and the amount
of informational asymmetry. The derivative of $\bar{u}_{1}$ and $\bar{u}_{2}$
with respect to $\gamma$ is%
\begin{equation}
\frac{\partial\bar{u}_{1}}{\partial\gamma}=\frac{\partial\bar{u}_{2}}%
{\partial\gamma}=\frac{e^{-2\gamma}k^{2}}{4}\left[  \frac{4e^{\gamma}}%
{(3\cosh\gamma+\sinh\gamma)^{3}}-s\right]  .\label{eq 4}%
\end{equation}
Eq. (\ref{eq 4}) shows that there is a threshold for the amount of
informational asymmetry, $s_{m}=4/27$. If $s>s_{m}$,%
\begin{equation}
\frac{\partial\bar{u}_{1}}{\partial\gamma}=\frac{\partial\bar{u}_{2}}%
{\partial\gamma}<0,
\end{equation}
which means the profits monotonously decrease as $\gamma$ increases. In this
case, the quantum game is definitely inferior to the classical game. It is
also interesting to see that if $s>1$ we can always find that for some value
of $\gamma$, $\bar{u}_{1}$ will be less than zero while $\bar{u}_{2}$ remains
positive. In this case, lacking information makes firm 1 lose money in
business on average, yet it is beyond firm 1's means to get out of it.

In the case that $s<s_{m}$,%
\begin{equation}
\left.  \frac{\partial\bar{u}_{1}}{\partial\gamma}\right\vert _{\gamma
=0}=\left.  \frac{\partial\bar{u}_{2}}{\partial\gamma}\right\vert _{\gamma
=0}>0,
\end{equation}
the profits increase as $\gamma$ increases when $\gamma$ is small. However we
can find $\gamma_{m}$ satisfying%
\begin{equation}
\left.  \frac{\partial\bar{u}_{1}}{\partial\gamma}\right\vert _{\gamma
=\gamma_{m}}=\left.  \frac{\partial\bar{u}_{2}}{\partial\gamma}\right\vert
_{\gamma=\gamma_{m}}=0,
\end{equation}
hence $\bar{u}_{1}$ and $\bar{u}_{2}$ simultaneously reach the maximum at
$\gamma=\gamma_{m}$. But when $\gamma>\gamma_{m}$ the profits decrease.

In the limit that $\gamma\rightarrow+\infty$, we have%
\begin{align}
\left.  \bar{u}_{1}(\gamma,s)\right\vert _{\gamma\rightarrow+\infty}  &
=\frac{k^{2}(1-s)}{8},\nonumber\\
\left.  \bar{u}_{2}(\gamma,s)\right\vert _{\gamma\rightarrow+\infty}  &
=\frac{k^{2}(1+s)}{8}.\label{eq inf}%
\end{align}
While in the classical game $\gamma=0$,%
\begin{align}
\bar{u}_{1}(0,s)  & =\frac{k^{2}}{9},\nonumber\\
\bar{u}_{1}(0,s)  & =k^{2}(\frac{1}{9}+\frac{s}{4}),
\end{align}
therefore if $1/9<s<s_{m}$, we can find $\gamma_{c}>0$ satisfying%
\begin{align}
\bar{u}_{1}(\gamma_{c},s)  & =\bar{u}_{1}(0,s),\nonumber\\
\bar{u}_{2}(\gamma_{c},s)  & =\bar{u}_{2}(0,s).
\end{align}
Thus we find another threshold for the amount of informational asymmetry,
$s_{c}=1/9<s_{m}$.

\begin{figure}
[b]
\begin{center}
\includegraphics[
height=5.1379cm,
width=8.1012cm
]%
{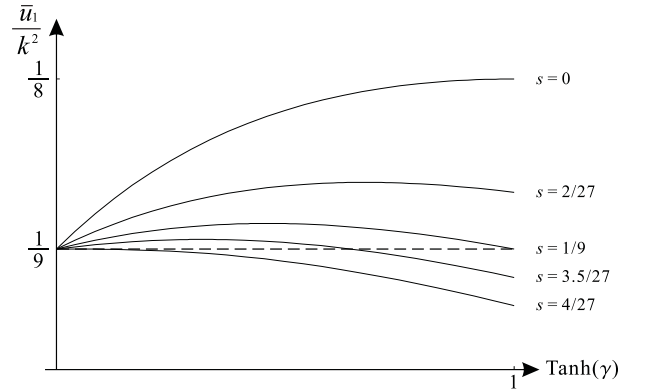}%
\caption{The $\bar{u}_{1}/k^{2}$ versus $\tanh(\gamma)$ plot with $k_{1}%
=k_{2}=k$. The solid lines are associated with the values of $s$. The
horizontal dashed line at $\bar{u}_{1}/k^{2}=1/9$ represents the classical
profit.}%
\label{Compare}%
\end{center}
\end{figure}

For $s<s_{c}$, $\left.  \bar{u}_{1}(\gamma,s)\right\vert _{\gamma
\rightarrow+\infty}>\bar{u}_{1}(0,s)$ and $\left.  \bar{u}_{2}(\gamma
,s)\right\vert _{\gamma\rightarrow+\infty}>\bar{u}_{2}(0,s)$, the quantum game
is always superior to the classical game for any $\gamma>0$. For
$s_{c}<s<s_{m}$, the quantum game is superior to the classical game for
$0<\gamma<\gamma_{c}$ but inferior for $\gamma>\gamma_{c}$, and the profits
reach the maximum at $\gamma=\gamma_{m}<\gamma_{c}$. While for $s>s_{m}$, the
quantum game is definitely inferior to the classical game, and the profits
will get worse when the entanglement increases. To be illustrative, we plot
firm 1's profit (divided by $k^{2}$) with different settings of $s$ in Fig.
\ref{Compare}, in which all the above intriguing features could be seen.%

In fact the profits in Eq. (\ref{eq 3}) consist of two parts: one is
independent of $s$, the other is linear with $s$. The first part is an
increasing function of $\gamma$ while the second is a decreasing one. The
combination of these two parts creates the intriguing features as mentioned
above. However it also implies that the quantum entanglement has contradictive
effects on the game with asymmetric information: on the one hand it
potentially increases the profits but on the other hand it potentially
decreases it. The part independent of $s$ in Eq. (\ref{eq 3}) can be regarded
as the representation of cooperation. As the entanglement increases, the
cooperation increases and the profits potentially increases. While the part
dependent of $s$ in Eq. (\ref{eq 3}) represents the impact of the
informational asymmetry. This impact will decrease the payoff with the
presence of entanglement, not only for the player who lacks information but
also for the one who possesses more information.

A special instance is the case with $s=0$ (see in Ref. \cite{6}), in which the
classical game turns back to the original one of symmetric information
proposed by Cournot \cite{7}. While in the maximally entangled limit with
$\gamma\rightarrow+\infty$, we have $\left.  \bar{u}_{1}(\gamma,0)\right\vert
_{\gamma\rightarrow+\infty}=\left.  \bar{u}_{2}(\gamma,0)\right\vert
_{\gamma\rightarrow+\infty}\rightarrow k^{2}/8$, which is exactly the Pareto
optimum. In this case the initial state tends towards the singular limit $%
{\textstyle\int}
\left\vert x,-x\right\rangle $d$x$. It is this limiting state, first
considered by Einstein, Podolsky and Rosen, which enables the two firms to
best cooperate, and therefore to be best rewarded. The dilemma-like situation
is thus completely removed in this limit.

\section{Conclusion}

We investigated the quantization of a game of a market situation known as the
Cournot's Duopoly of asymmetric information, based on the continuous-variable
model for quantum games given in Ref. \cite{6}. We found that, with the
presence of informational asymmetry, the quantum entanglement has
contradictive effects. On the one hand the quantum entanglement promotes
cooperation and potentially increases the profits. While on the other hand,
due to the asymmetric distribution of information, the quantum entanglement
induces a decreasing effect not only to the player who lacks information but
also to the one who possesses more information. The combination of these two
effects results in an intriguing variation of the game with respect to the
measure of entanglement and the amount of informational asymmetry.

\begin{acknowledgments}
We greatly appreciate Zeng-Bing Chen and Serge Massar for fruitful discussions
and valuable suggestions. This work was supported by the Nature Science
Foundation of China (Grant No. 10075041), the National Fundamental Research
Program (Grant No. 2001CB309300) and the ASTAR (Grant No. 012-104-0040).
\end{acknowledgments}

\end{document}